\documentstyle[12pt,twoside,fleqn,espcrc1]{article}
\newcommand{\mnote}[1]{\marginpar{\tiny { }}}   
\def\y0{y^{(0)}}

\title{Particle Production and Flow at SIS Energies }

\author{N.~Herrmann
\address{Gesellschaft f\"ur Schwerionenforschung, Darmstadt, Germany}
\address{Physikalisches Institut der Universit\"at Heidelberg, 
Heidelberg, Germany},
FOPI Collaboration
}

\begin{document}

\maketitle

\begin{abstract}
 An overview is given over recent measurement of flow and 
 particle production in the energy range from 0.1 to 2\,AGeV.
 Excitation functions for the directed sideward and the azimuthally symmetric 
 transverse flow are presented and show the importance of flow phenomena 
 in this incident energy regime.
 Rapidity density distributions are indicative of a system size dependence
 of the stopping process. 
 The role of strange particles as a probe for the hot and dense phase of 
 hadronic matter
 is discussed with respect to the production and propagation.
 The spectra of Kaons indicate an equilibration with the 
 surrounding baryons during the expansion while their 
 directed flow pattern is different from that of the nucleons. 
\end{abstract}

\section{Introduction}
Nuclear matter at densities of a few times the ground state density 
($\rho \leq 3 \rho_\circ$)
and at temperatures well below the Hagedorn temperature ($T \leq 100 MeV$)
is a system of strongly interacting hadrons whose bulk properties 
are still poorly understood.
The study of hadronic matter 
offers the possibility to test fundamental properties of QCD. 
Recent theoretical work suggests that effects of 
chiral symmetry restoration 
could give rise to dropping masses in the nuclear 
medium and should already be visible at relatively low 
densities \cite{KLI90,BRO91,HAT92,LI94,YAB94}.
In addition the hadronic matter state is the 
final state of any possibly produced quark-gluon-plasma state 
and should be reasonably well known in order 
to be able to detect a transition into a different phase. 
The properties of hadronic matter are, however, not easily accessible. 
Hadronic matter can only be produced in relativistic heavy ion collisions 
and the theoretical analysis based on transport equations is complex 
and has to take into account the dynamical evolution as well as 
the elementary processes.
The outcome of the reaction is not only determined by the 
mean field, the so called  nuclear matter equation-of-state, 
but at the same time by the properties of the constituents that
might be excited and form resonance matter. 
The multitude of dependences makes it essentially
impossible to relate specific properties to a single experimentally
measurable observable. The predictive power of the theory needs to be tested
by comparing simultaneously to several independent experimental observables.
Among them strangeness production and propagation is now 
experimentally accessible and found great theoretical 
interest \cite{AIC85,KAP86,SCH94,LUT94,LI95}.
 
With the installation of second generation experiments 
(EOS \cite{RAI90}, FOPI\cite{GOB93}, KaoS\cite{SEN93}, TAPS\cite{NOV91}) 
following the pioneering work at the Bevalac \cite{STO86,STOE86,GUT89}
much more complete information has become available.
This paper tries to 
summarize the current status of global hadronic 
observables and hadronic probes, i.e. strange particles, 
in the energy range from 0.1 to 2 AGeV and is organized in the 
following way:
Section~\ref{sec:sideflow} 
describes the excitation function of sideward flow. 
In section~\ref{sec:transflow}
the degree of collective motion is estimated from the average kinetic energies
and the transverse momentum spectra.
Section~\ref{sec:stopping} presents the latest information on 
stopping.
In section~\ref{sec:prod} the production yields of strange mesons are 
presented including preliminary new data on the $\Phi$\,-\,meson.
Section~\ref{sec:strangeflow} contains the status about the directed 
sideward flow of strange particles. 
Finally in
section~\ref{sec:summary} a brief summary is given.

\begin{figure}
 \vspace{7.5cm}
 \includegraphics{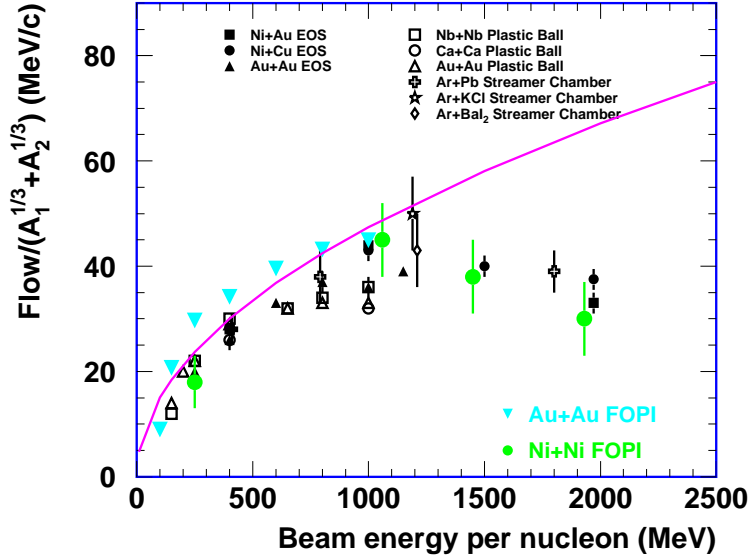}
 \caption{Excitation Function of Sideward Flow}
 \label{fig:eosexfl}
\end{figure}

\section{Sideward Flow }\label{sec:sideflow}
Sideward flow was proposed long time ago to carry the information about 
the nuclear matter equation of state \cite{STOE86}. Although it was later found 
that in addition Fermi momenta of the nucleons and 
two-body scattering processes drive the sideward flow \cite{BLA91}, 
experimentally it nevertheless remains interesting to measure the excitation 
function. 
The current status is depicted in fig.\ref{fig:eosexfl}: 
Plotted is the slope $F=d<p_x/A>/dy'$
of the average transverse momentum 
projected into the reaction plane with the normalized 
laboratory rapidity $y'=y/y_p$
for a variety of symmetric and asymmetric projectile\,-\,target combinations.
In order to make the different systems comparable to each other 
a scaling factor of  $(A_1^{1/3}+A_2^{1/3})$ as first used 
by J.Chance et al. \cite{CHA96} and suggested by Lang et al.\cite{LAN91} is 
applied. The data points originate mostly 
from the EOS collaboration \cite{CHA96} and are complemented by recent 
preliminary FOPI data \cite{CRO96}.
The events are selected according to the charged particle
multiplicity. They were chosen to cover an impact parameter range
as defined by the Plastic\,Ball collaboration 
with multiplicity bin M3+M4 \cite{DOS86}. The $F$\,-values shown in 
fig.\ref{fig:eosexfl} represents the flow of the average of $H$ and $He$ 
fragments only, although it is known that the heavier fragments 
shown an even more pronounced flow signature \cite{CRO96,PAR95}.

Within the current accuracy of comparison, namely the different acceptances 
for the global multiplicity and the different particle identification 
capability of the various experiments, the data are consistent.
A clear trend is visible: The sideward
flow is rising in the energy range from 150\,AMeV to 1\,AGeV according to the 
beam momentum in the CMS. This dependence is depicted by the solid line 
in Fig.~\ref{fig:eosexfl}. 
Beyond 1 - 1.2 AGeV the incident energy scaling is 
broken and the measured values are rather constant or even slightly 
dropping.
The mechanism that is responsible for this behaviour is not yet identified.
A systematic comparisons with dynamical models will have to reveal
whether the anisotropy of the NN-interaction, the 
excitation of resonances  and / or changes in the stiffness of the 
equation-of-state are responsibly for the observation. For this task
systematic errors can be reduced by applying the proper filter programs 
of the various experiments. 
 
\section{Transverse Flow }\label{sec:transflow}

\begin{figure}
\begin{minipage}[t]{7cm}
 \vspace{9.5cm}
 \includegraphics{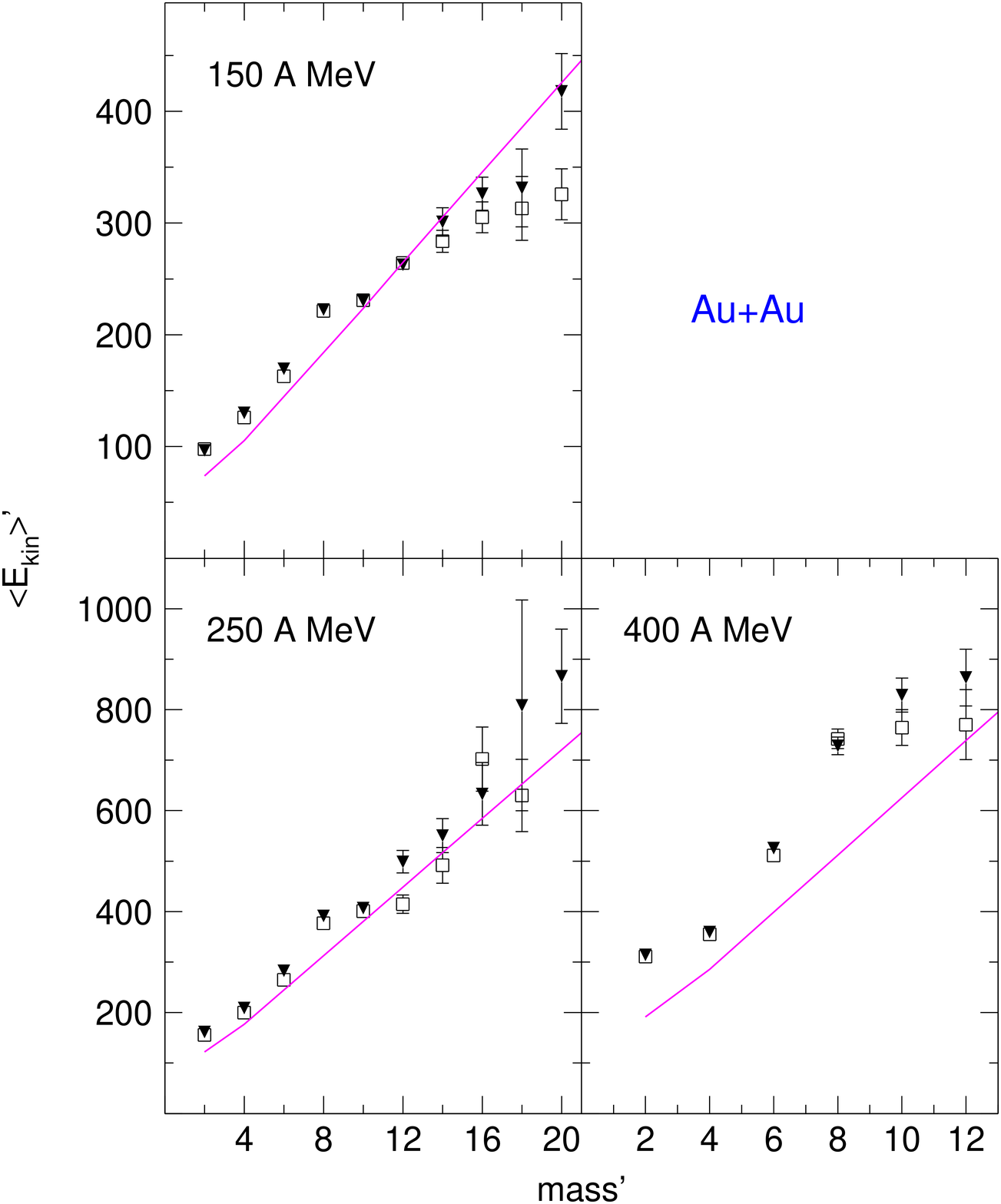}
\end{minipage} \ \hfill  \
\begin{minipage}[t]{7cm}
 \caption{Mean Kinetic Energies of Intermediate Mass Fragments}
 \label{fig:wrradfl}
 Mean kinetic energies of fragments with masses determined 
 by $A=2Z$ in the angular range $25^\circ < \Theta_CM < 45^\circ $
 are shown 
 for central collisions \cite{REI96}.
 Squares represent the data for an event samples selected by means of the 
 ratio of transverse to longitudinal kinetic energy (ERAT) of  200\,mb and 
 azimuthal symmetry,  triangles correspond to an ERAT selection of 50\,mb.
 The solid lines are given by blast model fits to the data.
\end{minipage}
\end{figure}

The energy contained in the sideward flow is only a small fraction 
of the available energy and thus does not influence the overall 
conditions and the thermalisation. The dominating collective flow 
at incident energies from 0.1 to 1 AGeV is an azimuthally symmetric flow 
component that can be recognized from its 
fragment mass dependence for the most central collisions \cite{JEO94}.
An example of this effect is presented in fig.\ref{fig:wrradfl} for 
the system Au+Au at 150, 250 and 400 AMeV \cite{REI96}.
An almost linear dependence of the average kinetic energies with 
the ejectile mass is observed. Such a behaviour is indicative 
for collective flow since under the assumption of a common temperature 
and a common flow velocity distribution at freeze-out, 
the average kinetic energy can be written as
$ <E_{kin}> = E_{thermal} + A*e_{collective} $.
The slope of the curves in fig.\ref{fig:wrradfl} determine 
$e_{collective}$ in a model independent way. The values are 
astonishingly high: 60\% of the available energy is found in the 
collective expansion. This large fraction of kinetic energy helps 
to understand the large abundance of intermediate mass fragments 
that indicate a small entropy, although a quantitative description 
by statistical models has not been achieved so far \cite{REI96}.
 
\begin{figure}
\begin{minipage}[t]{6cm}
 \vspace{11.cm}
 \includegraphics{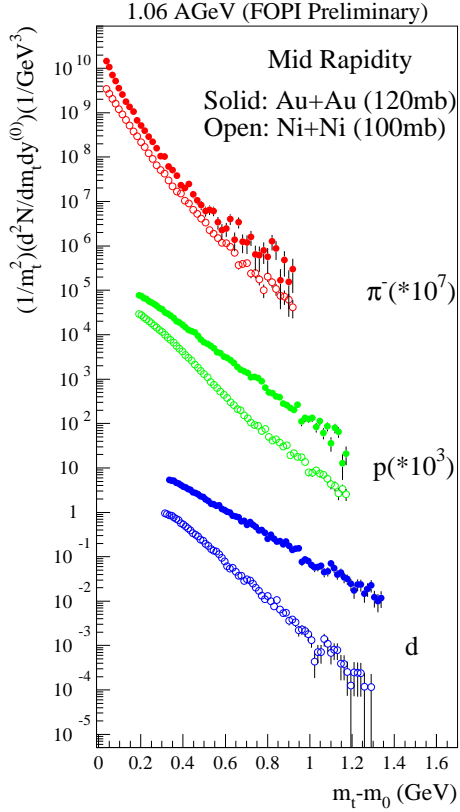}
\end{minipage} \ \hfill  \
\begin{minipage}[t]{8cm}
\caption{
Transverse mass spectra at midrapidity for Pions, Protons and Deuterons 
 in Ni+Ni and Au+Au at 1\,AGeV
}
\label{fig:mtauni}
Pion spectra show a concave shape whereas those of 
baryons can be described reasonably 
well by a single exponential function. The comparison of the 
spectra from the two systems reveals a system size dependence of the
slope parameters for the baryons.
\end{minipage}
\end{figure}

For higher incident energies the long lever arm offered by the IMF emission 
is lost for central collisions. 
Beyond 1\,AGeV the analysis is limited up to now 
to Hydrogen and Helium isotopes. 
Typical transverse mass spectra at midrapidity with a weighting 
factor of $1/m_t^2$ such that 
a thermal Boltzmann-like spectrum is represented by a straight line 
are shown in fig.\ref{fig:mtauni}. 
Event samples of 120\,mb ($b_{geo}=2fm$) and 100\,mb ($b_{geo}=1.8fm$) 
on the basis of charged particle multiplicity were used for 
the Au+Au and Ni+Ni system, respectively. 
For this type of analysis the centrality selection is not as
crucial any more, the results are already stable for the most central
400\,mb ($b_{geo}=3.6fm$). 
The proton and deuteron spectra can be described reasonably well 
by single exponential 
functions, while for the pion spectra the sum of two exponential 
functions is needed in order to describe the data.
Comparing the slopes of the different ejectiles for the same 
reaction one observes an ordering according to the ejectile mass:
$\pi$\,-\,specta are steeper than the proton spectra that themselves 
are exceeded by the deuteron spectra. 
It is also interesting to observe that at the same incident energy (1AGeV) 
the spectra change differently for different system sizes (Au+Au versus
Ni+Ni). The variation of the slope parameter is larger for the heavier 
system. 

\begin{figure}
 \vspace{8.5cm}
 \includegraphics{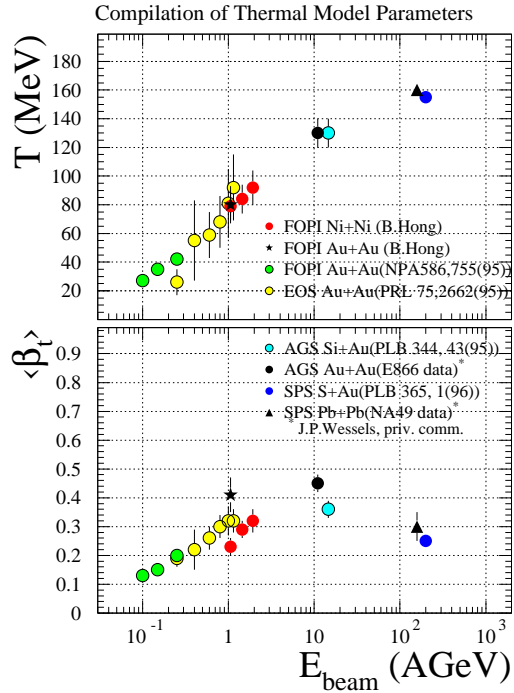}
 \caption{Excitation Function of Blast Model Parameters, Temperature $T$ 
and average transverse expansion velocity $\beta_t$}
 \label{fig:exradfl}
\end{figure}

The dependence of the slope parameter on the ejectile mass again is 
indicative for collective motion and 
motivates fitting the transverse mass distribution with a 
blast model hypothesis \cite{SIE79}. This approach was used lately by several 
authors. The summary of the available data in the energy range from
0.15 to 160\,AGeV where the mass dependence of the slope parameters
for the different particle species was used in order to extract the 
transverse flow component is shown in 
fig.\ref{fig:exradfl}. Preliminary FOPI data \cite{HON96} are shown in 
comparison to data from EOS \cite{LIS95} and AGS\cite{BRA95}, SPS\cite{BRA96} 
including preliminary analysis of the heavy systems \cite{WES96}.
Plotted are the average transverse velocities and the accompanying 
temperatures that are obtained as a second parameter from the fits.
It was checked that the results do not depend on the velocity
profile, e.g. using a linear dependence of the flow velocity with the 
radius gave the same result as using a fixed expansion velocity as suggested 
in \cite{SIE79}.
A clear trend is emerging from fig.\ref{fig:exradfl}: 
the temperatures are logarithmically 
rising over the full incident energy range from 0.1 to 160 AGeV and 
do not show a system size dependence. 
The average transverse expansion velocities 
are rising from 0.1 to 2 AGeV and show a system size dependence at 1\,AGeV
(see fig.\ref{fig:mtauni}).
A similar feature is observed at the higher incident energies when 
comparing the different systems \cite{WES96}.
At 2\,AGeV one observes collective expansion velocity values 
that are close to the ones obtained at an incident energy of 10.7 AGeV. 
The average transverse velocities seem to be limited to $<\beta_t> \leq 0.5$.
Extrapolating from fig.\ref{fig:exradfl} the maximum should be located 
at an incident energy of around $E_{beam}=5-10\,AGeV$. 

It should be mentioned at this point that the analysis of the heavier 
fragments (Fig.\ref{fig:wrradfl}) gives higher values for the average 
expansion velocities and accordingly lower temperatures. 
In trying to describe the distributions of heavier fragments with
nuclear charges $2\leq Z \leq 8$ with an isotropically expanding 
source \cite{REI96}, already at 400\,AMeV an average expansion velocity
of $<\beta> = 0.33$ is reached. 
This conclusion is in agreement with the light particle 
data \cite{POG95} that are included in fig.\ref{fig:exradfl}
and give a smaller value when analyzed for themselves. 
Their distributions
are probably influenced by evaporation so that the mass dependence
that gives rise to the collective flow estimate could be disturbed.
Interestingly enough the spectra
of the heaviest fragments show even a sensitivity to the 
flow velocity profiles at freeze-out.
The exact shape of the excitation function on radial flow 
thus depends on the knowledge of the distributions 
of different particle species with a large mass lever
asking for the future for an as complete 
measurement as possible with a special emphasis on the heavier fragments.

\section{Stopping}\label{sec:stopping}

\begin{figure}
 \vspace{7cm}
 \includegraphics{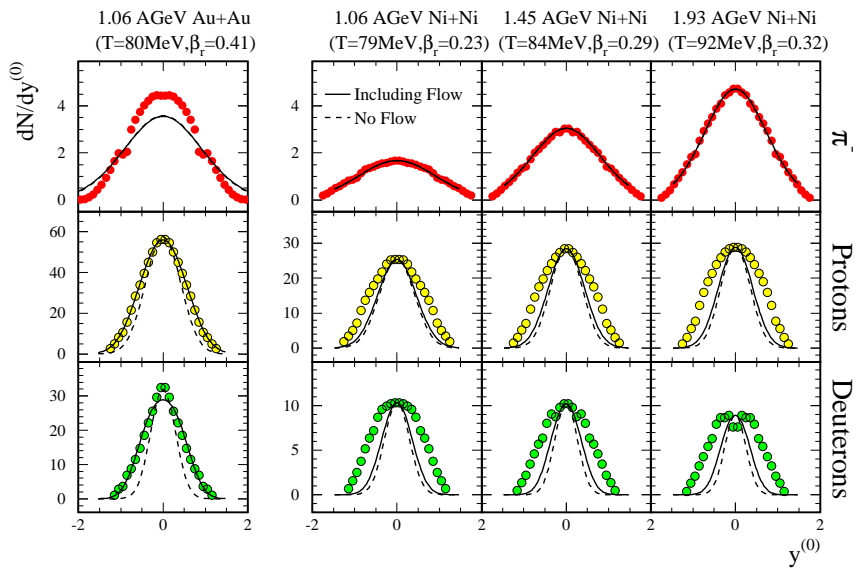}
 \caption{Rapidity Density Distributions}
 \label{fig:dndyauni}
\end{figure}

Once a substantial part of the populated phase space is measured by
the detector system, the question of nuclear stopping can be addressed.
For the lower incident energies $E_{beam} \leq 0.4\,AGeV$ and the heavy system 
an isotropically radiating source can be identified \cite{REI96}.
For the energy range above 1\,AGeV the situation is not so clear yet.
Information can be extracted from the 
exponential fits to the transverse mass spectra like in Fig.\ref{fig:mtauni} 
that allow to estimate the total yield of the 
emitted particles by integrating the fit functions from 0 to infinity.
The preliminary result of such an integration are shown in 
fig.\ref{fig:dndyauni} for the system Au+Au at 1AGeV and for the 
system Ni+Ni at 1, 1.45 and 1.93 AGeV.
The data are plotted versus the normalized center-of-mass rapidity 
$y^{(0)} = y/y_{projectile}$ and since for a symmetric
system the forward and backward hemisphere in the CMS have to be the same 
are symmetrized around midrapidity.
The data are compared to the predictions of an isotropic thermal model scenario
with (solid lines) and without (dashed lines) radial flow.  The parameters
used for the different systems are the ones shown in fig.\ref{fig:exradfl}
and are given in the figure.

Several interesting features can be noticed:
The distributions of protons and deuterons for Au+Au at 1\,AGeV are 
well described 
by an isotropically expanding source (with the parameter of the 
source determined by the transverse mass spectra at midrapidity). 
This Ansatz fails for the Ni+Ni 
system: a rapidity density distribution is observed that is wider 
than the one expected for an isotropically emitting source. 
While the data shown in fig.\ref{fig:dndyauni} correspond 
to an integrated cross section of 100\,mb this situation does not 
change even when selecting cross section as small as 30\,mb.
The enhanced longitudinal pattern is observed for the Ni system 
from 1 to 2 AGeV and indicates incomplete stopping. 
For the produced particles, e.g. pions, the deviations to the 
isotropic scenario are different: the data obtained for the different 
Ni measurements are compatible with the assumption of isotropic emission, 
while the pion distributions for the heavy system 
at 1 AGeV appear narrower than expected.

\section{Production of Strange Particles}\label{sec:prod}

\begin{figure}
 \vspace{7cm}
 \includegraphics{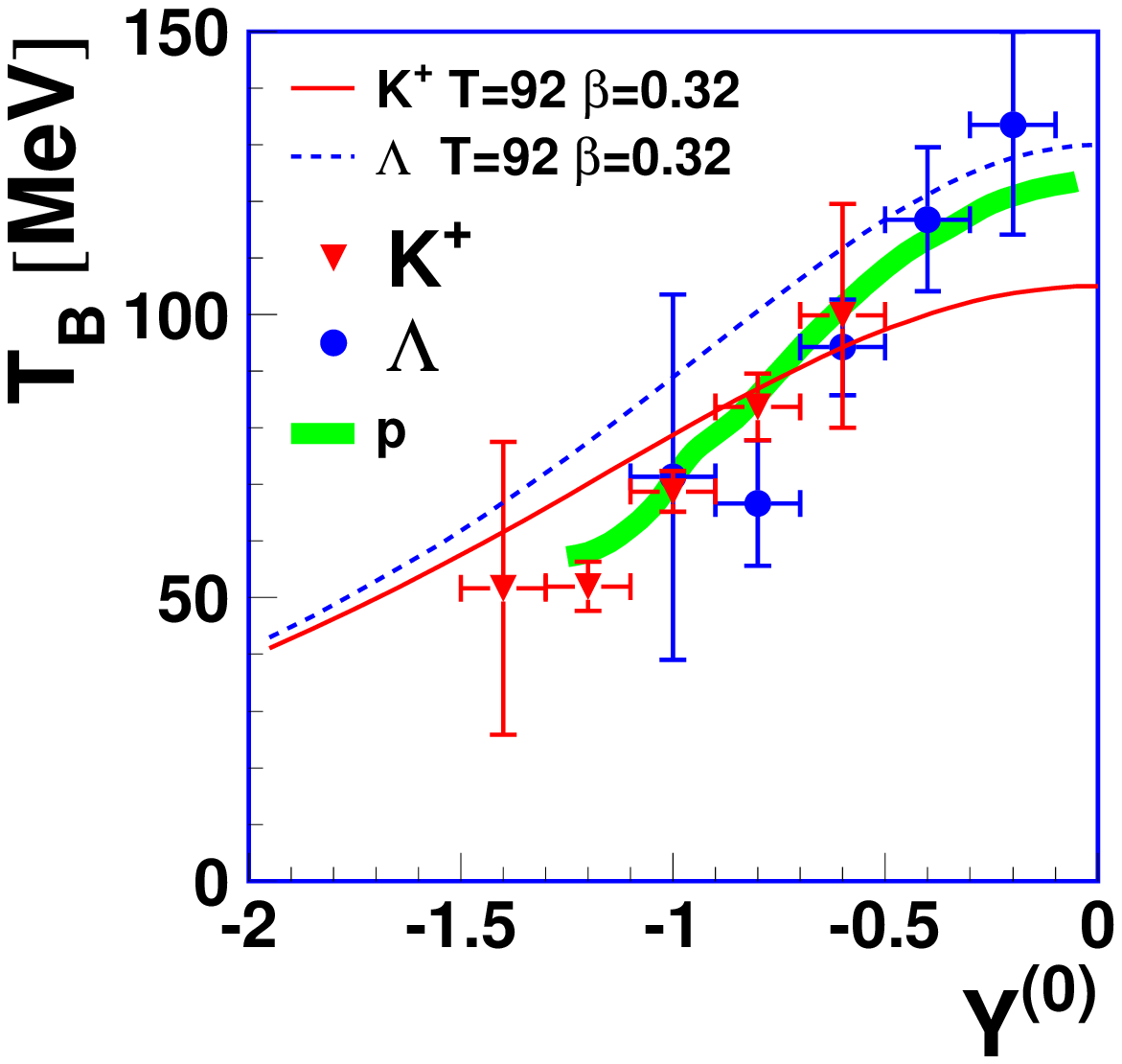}
 \includegraphics{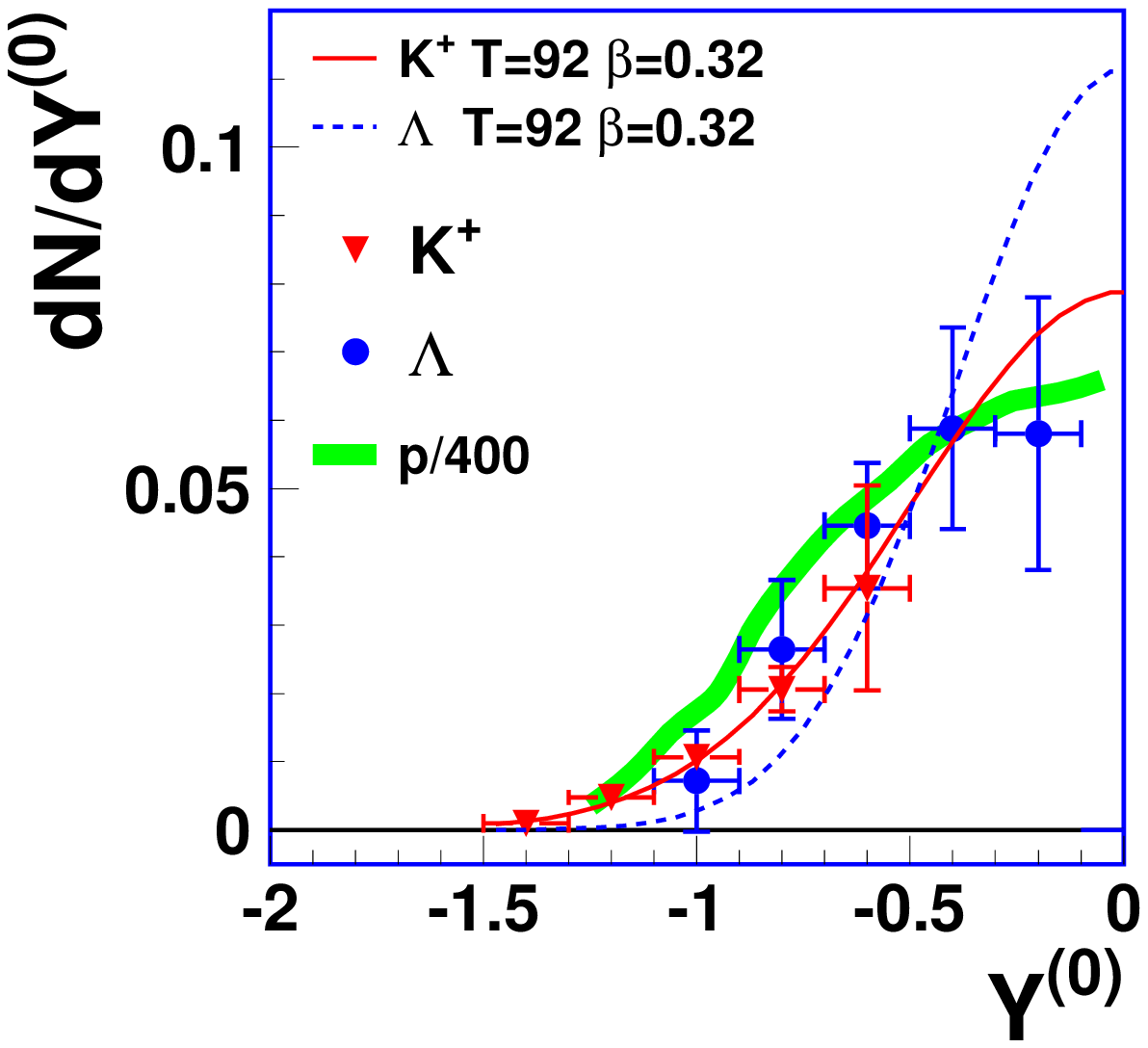}
 \caption{Slope Parameter and Rapidity Distributions of Strange Particles}
 \label{fig:strange}
\end{figure}

Since strangeness is conserved in strong interactions
strange particles are expected to carry a more direct signal 
of the hot and dense state that is eventually reached in the course 
of the collision. Especially the $K^+$ with its long mean free path 
could serve as a messenger.
Full phase space distribution are necessary in 
order to support this claim. Lately, the FOPI collaboration succeeded
to identify strange particles within a $4\pi$\,-\,detector at 
relative abundance of $1 \cdot 10^{-4}$ with respect to all particles 
\cite{RIT95}. Charged kaons are identified from a measurement of their 
specific energy loss in a central drift chamber, the curvature of the 
tracks and Time-of-flight of a scintillator barrel.
Neutral strange particles are reconstructed from tracks that do not 
originate from the primary vertex and are identified from their
invariant masses.
The particle identification capability allows to discuss the 
distributions of strange particles ($K^+$
and $\Lambda$) directly in comparison to the baryon distributions.
In order to bypass threshold and detection efficiency effects
transverse mass spectra were fitted by exponential functions 
and slope parameter and yields are obtained.
A comparison of the extracted parameters is shown in fig.\ref{fig:strange}.
Slope parameter (left panel) and integrated rapidity density distributions
(right side) are shown for Protons, Kaons and Lambdas. Error bars reflect
statistical errors only. A direct comparison to the protons show that
the strange particle distributions are very similar a) in the slope
parameter and b) with respect to the shape of the rapidity distributions.
Both observations favour the claim for a kinetic equilibrium and/or
substantial rescattering of the Kaons as well as of the Lambdas. 
On the other hand, within the framework of the expanding blast 
model scenario discussed so far, systematic differences are expected.
The predictions for an isotropically emitting expanding source are given by the
lines in fig.\ref{fig:strange}, depicting the typical cosh(y)\,-\,behaviour
for the slope parameters and almost gaussian shapes for the rapidity
distributions.
The $K^+$\,-\,data are reasonably well described by the assumption 
of a radially expanding source while the $\Lambda$\,-\,data show
systematic deviations that coincide with those observed for the protons.
The slopes for rapidities $y^{(0)} < -0.5$ are smaller than expected  
whereas the rapidity distributions are more elongated 
(compare fig.\ref{fig:dndyauni}).


\begin{figure}
\begin{minipage}[t]{7cm}
 \vspace{8.6cm}
 \includegraphics{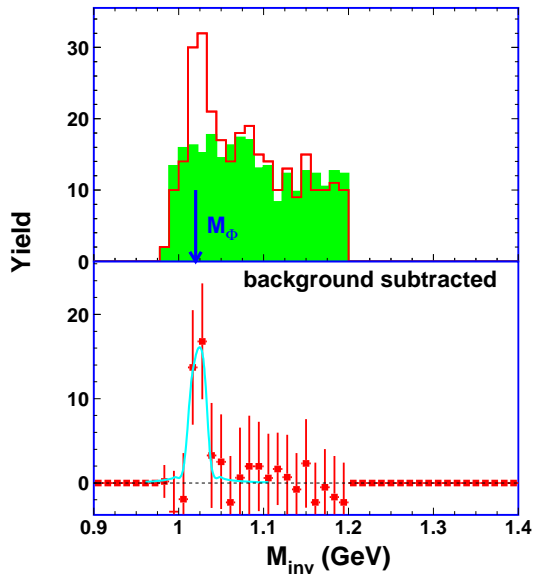}
\end{minipage} \ \hfill  \
\begin{minipage}[t]{7cm}
 \caption{Invariant Mass Distribution of K$^+$-K$^-$ pairs}
 \label{fig:phicand}
 The top panel shows the measured distribution as well as the one 
obtained from a mixed event analysis. The lower panel is obtained 
after subtraction of the uncorrelated background.
The spectrum is artificially cut off at 1.2 GeV.
\end{minipage}
\end{figure}

The phase space coverage of the FOPI\,-\,detector with
simultaneous identification of all charged particles also allows
to search for more exotic ones. 
In the context of strangeness production the most interesting one 
is the $\Phi$\,-\,meson.  
This resonance can be reconstructed from 
its decay $\Phi \rightarrow K^+ + K^-$ with a branching 
ratio of 49.1\%. Due to the narrow intrinsic width of $\Gamma = 4.4 MeV$ 
it can be recognized in the invariant mass distribution of $K^+-K^-$ candidate
pairs.
The reconstructed invariant mass distribution is shown 
in fig.\ref{fig:phicand} and shows a statistically significant 
enhancement at $M=1.020\,GeV$ the mass of the free resonance.
The combinatorial background was obtained from a mixed event analysis 
that employed the same cuts that were applied to the data.
A total of $30 \pm 8$ $\Phi$\,-\,mesons could be reconstructed in the 
reaction Ni+Ni at 1.93AGeV from an event sample of $7 \cdot 10^6$ events.
This is certainly not sufficient to make any claims on their 
distributions but it is enough to get an estimate on the 
production yield and points to interesting options for the future.
 
In order to arrive at a meaningful comparison of the production 
rates of the different particles species one has to extrapolate to the 
full solid angle. For the lighter $\pi$ and $K$\,-\,mesons an
extrapolation prescription based on the blast model predictions
can be justified by figs.\ref{fig:dndyauni} and \ref{fig:strange}.
For the $\Phi$ the same assumption of the emission from an
isotropically emitting expanding source was used in order to determine 
the detection efficiency. The source parameter were determined from 
the baryon transverse mass spectra at midrapidity ($\beta=0.32,T=92MeV$).
The overall detection efficiency  was estimated from a MonteCarlo
simulation of the complete detector response, tracking and 
identifying scheme starting 
with a thermal $\Phi$\,-\,distributions with the parameters given above.
Clearly much more statistics is needed to verify the assumptions and 
narrow down the systematic errors that are estimated to be smaller
than a factor of 2, when neglecting the influence of the unknown 
angular distribution. 
Sensitivity to changes of the mean and the width 
of the peak requires a substantial increase in the statistics and
thus dedicated running time.
    
\begin{figure}
 \vspace{7cm}
 \includegraphics{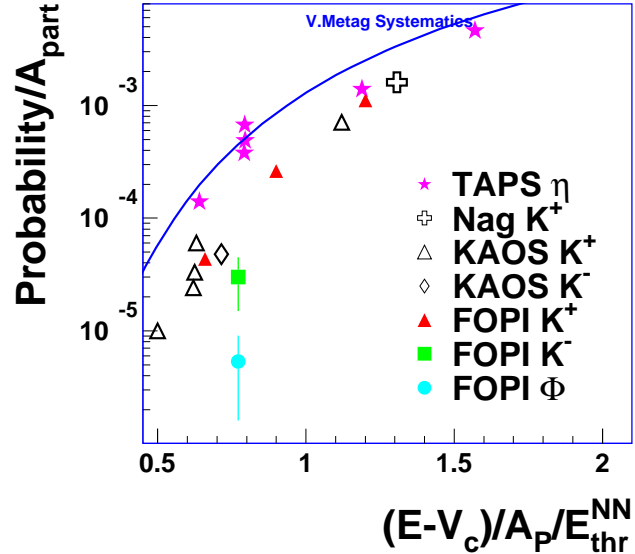}
 \caption{Comparison of Extrapolated Production Probabilities of Strange Mesons}
 \label{fig:strangeprod}
\end{figure}

Despite the difficulties discussed above it is instructive to 
compare the production yield for the different strange mesons.
This is done in fig.\ref{fig:strangeprod} where 
in addition to the directly observed Kaons and Antikaons 
and the estimate for the  $\Phi$ of FOPI,  recent data on $K^+$ and 
$K^-$ from KaoS \cite{SEN96} and on $\eta$ from TAPS \cite{AVE96} 
are included.
The production probabilities per participant nucleon are plotted
versus the incident energy above the Coulomb barrier normalized 
by the threshold energy that is necessary to produce a certain
particle in a NN\,-\,collision. This representation was introduced 
by Metag \cite{MET93} and yields a very consistent description for 
all the pion data from 0.02AMeV up to 2AGeV including the FOPI 
pions: it results in the solid line of fig.\ref{fig:strangeprod}. 

Some preliminary conclusions can be drawn from 
inspection of fig.\ref{fig:strangeprod}:
\begin{itemize}
\item The production probabilities extrapolated from the FOPI data 
are consistent with those obtained at midrapidity with a dedicated 
spectrometer (KaoS \cite{SEN96}).
\item The strangeness degree of freedom is not equilibrated, e.g. 
strange particles are not produced with a weight given by the mass 
relative to the corresponding threshold energy.
\item The incident energy dependence seems to be stronger as 
compared to the pion systematics (given by the solid line).
\item $K^-$ and $K^+$ yield are comparable at the same available energy.
This is surprising since the $K^-$- production is suppressed in pp collisions 
by an order of magnitude relative to $K^+$- production 
at the same available energy \cite{SEN96}, e.g. when taking into account 
the different production thresholds in 
$pp \rightarrow nK^+\Lambda$ and $pp \rightarrow ppK^+K^-$.
\item $\Phi$\,-\,production occurs at a level of $10^{-2}$ to pions 
and 10\% with respect to $K^-$. The later number is similar to observations 
made at AGS energies \cite{COL95} and could be indicative for a coalescence
like production mode.
\end{itemize}

The production probability systematics stresses the role of 
strange particles as a interesting probe for the properties 
of hot and dense hadronic matter. 
It should be noted that especially the observed production rate
of $K^-$ represents a puzzle since a) the elementary production rate is lower
and b) the absorption is stronger as compared to $K^+$ \cite{SEN96}.
Whether these losses can be balanced by additional production 
channels, e.g. $\Lambda \pi \rightarrow K^-N$, is questionable.   
In addition the phase space description fails to reproduce the 
various mass systems at the same incident energy.
This deficiency is caused by a more than linear increase of the 
Kaon production yields with the number of participating nucleons 
\cite{SEN96,MIS94}, clearly indicating that central collisions 
are not just a superposition of independent NN\,-\,collisions.

A consistent model to describe all the available particle yields is 
not available yet. Most of the analyses have focused on the $K^+$
distributions so far. The comparison with transport models shows
that $N\Delta$ and $\Delta$$\Delta$ collisions are necessary in order 
to account for the observed Kaon yield \cite{FAN94,HAR94}.
According to those models Kaons are predominantly produced in the 
early stages of the reaction at high temperature and densities.
Since the elementary production cross sections are poorly known
the uncertaintities in the quantitative conclusions are fairly 
large. The observed rates are, however, so large that 
a soft equation of state is needed in order to provide a sufficient 
number of collisions.

Whether one needs additional in-medium effect like the 
modification of the particle masses in order to explain the $K^+$ 
yields (like in \cite{FAN94}) is not clear at this moment. 
Modification of the (Anti)Kaon
masses as they are expected from chiral perturbation theory 
could offer an explanation of the observed $K^-$ yield. 
All the theoretical attempts dealing with in medium Kaon masses 
\mnote{check refs!}
\cite{SCH94,LUT94,FAN94} predict a slightly rising Kaon and a 
more strongly dropping 
Antikaon mass when the baryon density is increased. $K^-$\,-\,mesons 
are therefore much more easily produced in dense matter.

Clearly those
ideas need to be tested by independent observables. 
Within the framework of chiral perturbation theory dropping 
in-medium masses are caused by scalar and vector potentials that 
also influence the propagation of the particles 
through the matter \cite{LI95}.
It is therefore very important to try to measure the 
flow of the strange particles with respect to the baryon flow.

\section{Strangeness Flow}\label{sec:strangeflow}

\begin{figure}
\begin{minipage}[t]{7cm}
 \vspace{9.5cm}
 \includegraphics{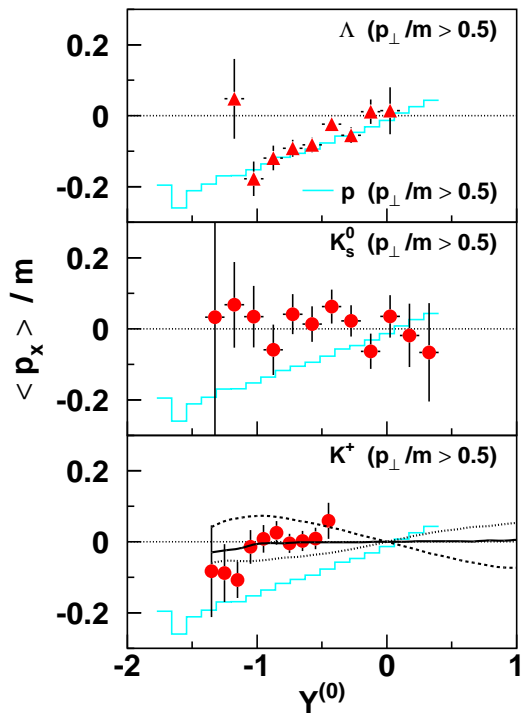}
\end{minipage} \ \hfill  \
\begin{minipage}[t]{7cm}
 \caption{Sideward Flow of Strange Particles}
 \label{fig:sflow}
 Average transverse momenta per mass projected onto the reaction 
 plane are shown as function of the normalized rapidity $\y0$
 for $\Lambda$ (top), $K^\circ$(middle) and $K^+$ (bottom) in comparison
to protons (histogram). The $K^+$ distributions are compared to the 
predictions from an RBUU calculation of Li and Ko \cite{LI96} that 
make use of different in-medium potentials: 
the dashed dotted, dashed and solid  lines represents the cases
{\it no potential, vector potential only and scalar+vector potential},
respectively.
\end{minipage} 
\end{figure}

First results on the directed sideward flow of strange particles 
have become available recently \cite{RIT95}. For the Ni+Ni reaction
at 1.93 AGeV it is possible to determine the average in\,-\,plane 
momentum $<p_x>$ 
of Protons,  Lambdas, $K^+$ and $K^0$ under the same kinematical cuts.
The experimentally measured values  
are shown in fig.\ref{fig:sflow} for a $p_t/m$\,-\,cut of 0.5 as function
of the rapidity. The events used for this comparison were selected 
by means of  
charged particle multiplicity  and represent the most central ones with 
an integrated cross section of 420\,mb. The reaction plane resolution 
was determined to $\Delta \phi = 40^\circ$.
Note that the momentum cut enhances the $<p_x>$ value by more than a factor of 
2 at target rapidity.

In all three panels protons are represented by the solid histograms.
In the upper panel they are compared to $\Lambda$\,-\,particles.
The distributions agree within the error bars and support the 
earlier observation that the phase space distributions of protons 
and Lambdas are very much alike.
The middle and the lower panel of 
fig~\ref{fig:sflow} shows the same comparison to the protons 
to $K^0_s$ and $K^+$, respectively. For both particle species no sideward flow
signal is observed, e.g. the emission of those Kaons is 
independent of the orientation to the reaction plane. 
Since the production of Kaons and Lambdas at incident energies below
2\,AGeV proceeds in an associate manner ($ NN \rightarrow NK\Lambda $ or
$N\Delta \rightarrow NK\Lambda $) the observed differences 
have to be attributed
to the propagation process through the expanding nuclear medium.
$\Lambda$s seem to be attracted by the baryons whereas $K^+$ and $K^0$ 
are repelled.

The vanishing $K^+$ sideward flow was predicted by the RBUU transport model 
of Li and Ko under the assumption that both a scalar and a vector potential
act on Kaons \cite{LI95,LI96}. 
As can be seen from the bottom panel of Fig.\ref{fig:sflow}, where
the model predictions folded with the experimental resolution are shown
for various options of the Kaon potential,
the data are incompatible with the assumption of free propagation
of Kaons. According to the RBUU calculation the non-interacting Kaons 
should exhibit a small in\,-\,plane flow in the direction of the 
baryons. The origin of this sideward flow was traced to the 
production kinematics in $NN$ and $N\Delta$ collisions \cite{LI96}.
The Kaons thus have to experience a force that repels them from 
the baryons.  The comparison with the data shows that the vector 
potential is responsible for the repulsion, but it is too strong
when it is not balanced at the same time by the scalar potential.
The in-medium potentials also offer an explanation for the 
$\Lambda$\,-\,flow. In this case the potential is attractive 
and the Lambdas are pulled into the regions of high baryon density
and pick up the relatively strong baryon flow \cite{LI96}.
Before taking the agreement as a proof for in-medium potentials it 
has to be mentioned there are indications
that using a different transport code even without 
invoking in-medium potentials, the experimental findings
can be described \cite{AIC96}.
Clearly, a consistent comparison of the full event information 
is necessary before drawing final conclusions.  An additional crucial test 
will be the extension of the 
current signals to Antikaons and the impact parameter dependence of 
the Kaon flow that will become available soon.

\section{Summary and Conclusions}\label{sec:summary}

The amount of data available on heavy ion reaction around 1\,AGeV 
has considerably improved and shows several remarkable features:
a) The baryonic sideward flow is rising from 0.1 to 1\,AGeV and 
seems to decrease or saturate above.
b) The systems seem to explode with expansion velocities that 
are increasing in the incident energy range from 0.1 to 2\,AGeV.
c) Full stopping is only achieved for the heaviest system.
d) Strangeness production in the near threshold region around 1-2\,AGeV
is not in chemical equilibrium, although  the spectra of 
the strange particles are 
consistent with the assumption of a kinetic equilibrium.
e) Differences between strange baryons and mesons are observed for the 
sideward flow.  
So far this full set of observations is not consistently accounted for by any
dynamical theory. Since for some of the single observations
like the $K^-$ production rate fundamental changes of the properties
of mesons in the nuclear medium seem to be necessary it remains 
a challenging task to compare all the available information to the 
predictions of theory consistently.
This is especially important since 
at those energies one might have the unique opportunity 
to look at fairly low temperatures and high baryon densities 
at the consequences of fundamental symmetries of QCD, namely the   
partial restoration of chiral symmetry. 

\section*{Acknowledgments}
I would like to thank all members of the FOPI collaboration of SIS/GSI:
B.~Hong$^{4}$, J.~Ritman$^{4}$, D.~Best$^{4}$,
A.~Gobbi$^{4}$, K.~D.~Hildenbrand$^{4}$, Y.~Leifels$^{4}$, 
 C.~Pin\-ken\-burg$^{4}$, W.~Reisdorf$^{4}$, D.~Sch\"ull$^{4}$,
U.~Sodan$^{4}$, G.~S.~Wang$^{4}$, T.~Wie\-nold$^{4}$, 
J.~P.~Alard$^{3}$, V.~Amou\-roux$^{3}$, N.~Bastid$^{3}$, I.~Belyaev$^{7}$,
L.~Berger$^{3}$, J.~Bie\-gan\-sky$^{5}$, A.~Buta$^{1}$, R.~\v{C}aplar$^{11}$,
N.~Cin\-dro$^{11}$, J.~P.~Coffin$^{9}$, P.~Crochet$^{9}$, R.~Dona$^{9}$,
P.~Dupieux$^{3}$, M.~Dzelalija$^{11}$, M.~Eskef$^{6}$, P.~Fintz$^{9}$, Z.~Fodor$^{2}$,
A.~Genoux-Lubain$^{3}$,  G.~Goebels$^{6}$, G.~Guil\-laume$^{9}$,
E.~H\"afele$^{6}$,  S.~H\"olb\-ling$^{11}$,
F.~Jundt$^{9}$, J.~Kecske\-me\-ti$^{2}$,
M.~Kirejczyk$^{4,10}$, Y.~Korchagin$^{7}$, R.~Kotte$^{5}$, C. Kuhn$^{9}$,
D.~Lam\-brecht$^{3}$, A.~Lebedev$^{7}$, 
A. Le\-be\-dev$^{8}$, I.~Legrand$^{1}$,
C.~Maazouzi$^{9}$, V.~Manko$^{8}$,
J.~M\"os\-ner$^{5}$, S.~Mohren$^{6}$, W.~Neubert$^{5}$,
D.~Pelte$^{6}$, M.~Petrovici$^{1}$, F.~Rami$^{9}$,
V.~Ramillien$^{3}$, C.~Roy$^{9}$, 
Z.~Seres$^{2}$, B.~Sikora$^{10}$, V.~Simion$^{1}$,
K.~Siwek-Wilczy\'{n}ska$^{10}$, V.~Smol\-yan\-kin$^{7}$, 
L.~Tizniti$^{9}$, M.~Trzaska$^{6}$, M.~A.~Vasiliev$^{8}$, P.~Wagner$^{9}$,
D.~Wohl\-farth$^{5}$ and A.~Zhilin$^{7}$\\
$^1$ IPNE, Bucharest,
$^2$ CRIP Budapest, 
$^3$ LPC Clermont-Ferrand,
$^4$ GSI, Darmstadt, 
$^5$ FZ Rossendorf, Dresden,
$^6$ Universit\"at Heidelberg, 
$^7$ ITEP Mos\-cow, 
$^8$ KI Moscow, 
$^9$ CRN and Universit\'{e} Strasbourg, 
$^{10}$  Warsaw University, 
$^{11}$ RBI Zagreb 


\end{document}